\documentclass[prb, twocolumn]{revtex4}
\usepackage[toc,page]{appendix}
\usepackage{amsfonts}
\usepackage{amsmath}
\usepackage{amssymb}
\usepackage{graphicx}
\usepackage{epstopdf}
\usepackage{color}
\usepackage{braket}
\usepackage{tabularx}
\DeclareMathOperator{\Tr}{Tr}

\input epsf.sty

\begin{document}

\title{Manifestation of chiral magnetic current in Floquet-Weyl semimetals}
\author{Tsung-Yu Chen$^{1}$, Po-Hao Chou$^{2}$ and  Chung-Yu Mou$^{1,3}$}
\affiliation{$^{1}$ Center for Quantum Science and Technology and Department of Physics, National Tsing Hua University, Hsinchu, Taiwan 300, R.O.C.}
\affiliation{$^{2}$ Physics Division, National Center for Theoretical Sciences, Taipei 10617, Taiwan, R.O.C.}
\affiliation{$^{3}$ Institute of Physics, Academia Sinica, Nankang, Taiwan, R.O.C.}

\begin{abstract}
Materials that can host macroscopic persistent current are important because they are useful for energy storage. However, there are very few examples of such materials in nature. Superconductors are known as an example in which flow of supercurrent can persist up to 100,000 years. The chiral magnetic current is possibly the second example predicted by the chiral magnetic effect.  It was
 proposed to be realized in recently discovered Weyl semimetals. However,  a no-go theorem negates the chiral magnetic effect and shows that the chiral magnetic current is generally absent in any equilibrium condensed-matter system. Here we show how to break the no-go theorem by resorting to dynamical transitions in time-frequency space.  By driving an insulator using a time-periodic potential and coupling it to a phonon heat bath that provides suitable dissipation, we show that a Floquet-Weyl semi-metallic phase with Fermi-Dirac-like distribution emerges. Furthermore, we show that even in the presence of a static magnetic field, the resulting steady Floquet-Weyl semimetal supports non-vanishing chiral magnetic current. Our dynamical model provides a systematic way to fully realize the chiral magnetic effect in condensed matter systems. 
\end{abstract}

\pacs{74.70.Xa, 74.20.Mn, 74.20.Rp}
\maketitle
\section{Introduction}
The flow of electric current in a medium is usually dissipative. Hence without supplying energy, the current  can not be maintained indefinitely.  However, there are a few cases in which quantum mechanics dictates that the equilibrium state of a medium can support a persistent current.  In the case of superconductors, a persistent super-current  can be set up in a superconducting loop with duration being over 100,000 years\cite{persist}. In magnetized materials, persistent microscopic currents always accompany the magnetization. However,
this type of current is a bound current and can not be extracted as a free current.  Similarly, tiny persistent currents can also arise in mesoscopic metallic rings that are placed in a magnetic field when the size of the metallic system is reduced to the scale of the electron coherence length\cite{persist_2}. 

On the other hand,  the discovery of graphene and subsequent uncovering of topological materials\cite{graphene,Kane2005_7,Zhang2012,Liang2008,Yan2017} have inspired the idea to realize macroscopic persistent current in materials that exhibit the chiral magnetic effect (CME).  The CME is a manifestation of the chiral anomaly and is believed to be able to host macroscopic persistent currents in systems composed of relativistic massless fermions\cite{CME}.  In condensed matter systems with topological phases,  relativistic massless quasi-particles would emerge in low energies\cite{Dirac1,mou1} of the semimetallic phase at the transition point when the system goes from topological non-trivial phase to topological trivial phase. In particular, in the presence of the time-reversal and inversion symmetries, the semimetallic phase is characterized as the 3D Dirac semimetal\cite{Dirac2, Dirac3, Dirac4} with the underlying quasiparticles being Dirac fermions without definite handedness (or chirality). When either the inversion symmetry or the time-reversal symmetry is broken, Weyl fermions with definite chirality can be also realized. \cite{Weyl1, Weyl2, Weyl3,Weyl4}

Due to the Nielsen-Nimomiya theorem\cite{Nielsen}, Weyl fermions with definite chirality in condensed matter systems on a lattice cannot exist alone. 
The net chirality due to all Weyl nodes must vanish. Hence Weyl nodes must occur in pairs with opposite chiralities. One of peculiar effects associated with Weyl nodes is the chiral anomaly\cite{Weyl4}, which states that even though the total chirality due to all Weyl nodes vanishes, the chiral current is non-conserved.  As a result, in the presence of a static magnetic field $\vec{B}$, a macroscopic persistent current, $\vec{J}$, in parallel to the magnetic field  is predicted to arise such that $\vec{J}=\alpha \vec{B}$ with $\alpha$ being the CME coefficient. $\alpha$ is proportional to the energy separation between Weyl nodes. In the continuum model of Weyl fermions, a non-vanishing $\alpha$, i.e., the CME effect is generally confirmed in theoretical calculations. However, in condensed matter systems where Weyl fermions are put on latices, a no-go theorem indicating that the chiral magnetic current must vanish is established\cite{nogo}. More precisely, the no-go theorem shows that for equilibrium condensed matter systems in static magnetic fields, the electric current can be expressed as the total derivative of band energy dispersion. As a result of the periodicity of energy dispersion in Brillouin zones, the ground state  does not support any persistent current\cite{nogo}. On the other hand,  derivations of $\alpha$ based on the Kubo formula indicate that only in ac magnetic fields, condensed matter systems may support the chiral magnetic current.\cite{w_01,w_02,w_03,w_04,w_05}
''
The derivation of $\alpha$ leads to the consideration of driving condensed matter systems by time-periodic perturbations. For condensed matter systems driven by time-periodic potentials or gauge potentials, additional Brillouin zones known as Floquet Brillouin zones are created along the time axis\cite{Floquet_1, Floquet_2}.  This additional periodicity modifies the band structure and the current operator. Furthermore, it can be used to create the so-called Floquet-Weyl semimetallic phase.  There has been extensive work on creating the Floquet-Weyl semimetallic phase via time-periodic potentials \cite{Floquet_top_1, Floquet_top_2} or gauge fields\cite{Floquet_top_3, Floquet_top_4} generated by irradiation of light.  The realization of the chiral anomaly in the Floquet-Weyl semimetallic phase or static Weyl semimetallic phase is also discussed. For instance, it is shown that anomalous charge density will be generated as a manifestation of the chiral anomaly\cite{Floquet_CME_1}. It is also shown that it is possible to generate large electric current by using the polarized light acting as an effective magnetic field in Weyl semimetals\cite{Floquet_CME_2}.  A single Weyl node is shown to be realizable in Floquet lattice systems\cite{Floquet_CME_3} so that the CME can be realized.  Furthermore, it is shown that the Nielsen-Ninomiya no-go theorem still holds on a Floquet lattice, but in the
adiabatic limit,  the time evolution of the low-energy section is decoupled from the high-energy section so that unbalanced handedness of Weyl fermions can be created to realize the CME\cite{Floquet_CME_4}. While these works show that the CME current could be realized in Floquet systems, the considerations are based on the lowest energy band of the Floquet system. Therefore, the  electric current can be expressed as total derivative of band energy dispersion and it may still suffer from the problem and issue governed by the no-go theorem.

In this work, we show that one can break the no-go theorem by including dynamical transitions involved with higher Floquet bands such that the current operator is no longer the derivative of the energy dispersion for quasi-particles. 
Furthermore, by coupling the Floquet-Weyl semimetallic band to a phonon heat bath that provides suitable dissipation, a steady state that features the Fermi-Dirac-like filling of the Floquet bands can be established. 
As a result, we show that even in the presence of a static magnetic field, the resulting steady Floquet-Weyl semimetallic phase supports non-vanishing chiral magnetic current, explicitly realizing the chiral magnetic effect in a condensed matter system. 

\section{Theoretical Model} 
We start by considering a 3D multi-layer system driven by a time-periodic potential governed by the Hamiltonian $H=\sum_{\mathbf{k}, \sigma, \sigma'= \uparrow \, {\rm or} \, \downarrow}  f^\dag_{\mathbf{k} \sigma} h^{\sigma \sigma'}_{\mathbf{k}} f_{\mathbf{k} \sigma'}$ with $h_{\mathbf{k}}=h_0({\mathbf{k}})+h_T({\mathbf{k}},t)$. Here
$h_0$ is the static part of $h_{\mathbf{k}}$ that governs the multi-layer system composed by 2D topological insulators and magnetic layers 
and is given by \cite{Burkov2011}
\begin{align}
    h_0 ({\mathbf{k}}) = &\sin k_x\sigma_x + \sin k_y\sigma_y \nonumber + \sin k_z \sigma_z\\
    & + ( m_0 - \cos k_x - \cos k_y)\sigma_z.
\end{align}
Here  the lattice constant is set to be $1$, $\mathbf{k}=(k_x,k_y,k_z)$ is the wave-vector, the first three terms describe the spin-orbit coupling with the coefficients being set to be one so that all energies in the followings are expressed in unit of the spin-orbit coupling, and the 4th term is due to the presence of magnetization $m_0$.  $h_T(h_{\mathbf{k}}, t)$ is a time-periodic driven potential and is given by 
\begin{align}
  h_T({\mathbf{k}}, t) =  s\cos k_z \bar{\theta} (t-\frac{T_0}{2})  + m_1 \cos(\Omega t)\sigma_z,
\end{align}
where $m_1$ is the amplitude for on-site oscillating magnetization along $\hat{z}$,  $s$ is the strength of a background periodic step-function driven field
$\bar{\theta} (t-\frac{T_0}{2}) $. Here $\bar{\theta} (t-\frac{T_0}{2})$ is defined by
\begin{equation}
\bar{\theta} (t-\frac{T_0}{2}) = \left\{ \begin{matrix}
    1, & t \ge T_0/2 + nT_0, \\
    0, & t < T_0/2 + nT_0, 
\end{matrix} \right.
\end{equation}
with $n$ being any given integer and $T_0$ being the period of the driving potential.  Note that because Weyl nodes appear at $k_z = 0$ or $\pi$, the value of $s$ 
affects the energy difference of Weyl nodes directly. More precisely, the first term in Eq. (2) is introduced to shift and control positions of Weyl nodes in $k_z$. The action of switching-on and -off  by the driven field
$\bar{\theta} (t-\frac{T_0}{2})$ is to control the effective amplitude of $\cos k_z$ and we choose the duration of switching-on to be half period so that in the effective Hamiltonian, the amplitude of $\cos k_z$ is 1/2. Physically, as $\cos k_z$ corresponds to the nearest-neighbour hopping along z-axis in real space, it corresponds to periodically turned-on hopping along z-axis and can be realized by using arbitrary waveform generator (AWG). Similarly, the second term - cos-like driving of the magnetization can be realized either by using AWG or using ac magnetic fields to generate  oscillating magnetization. 

\begin{figure}[t]
\includegraphics[width = 9cm]{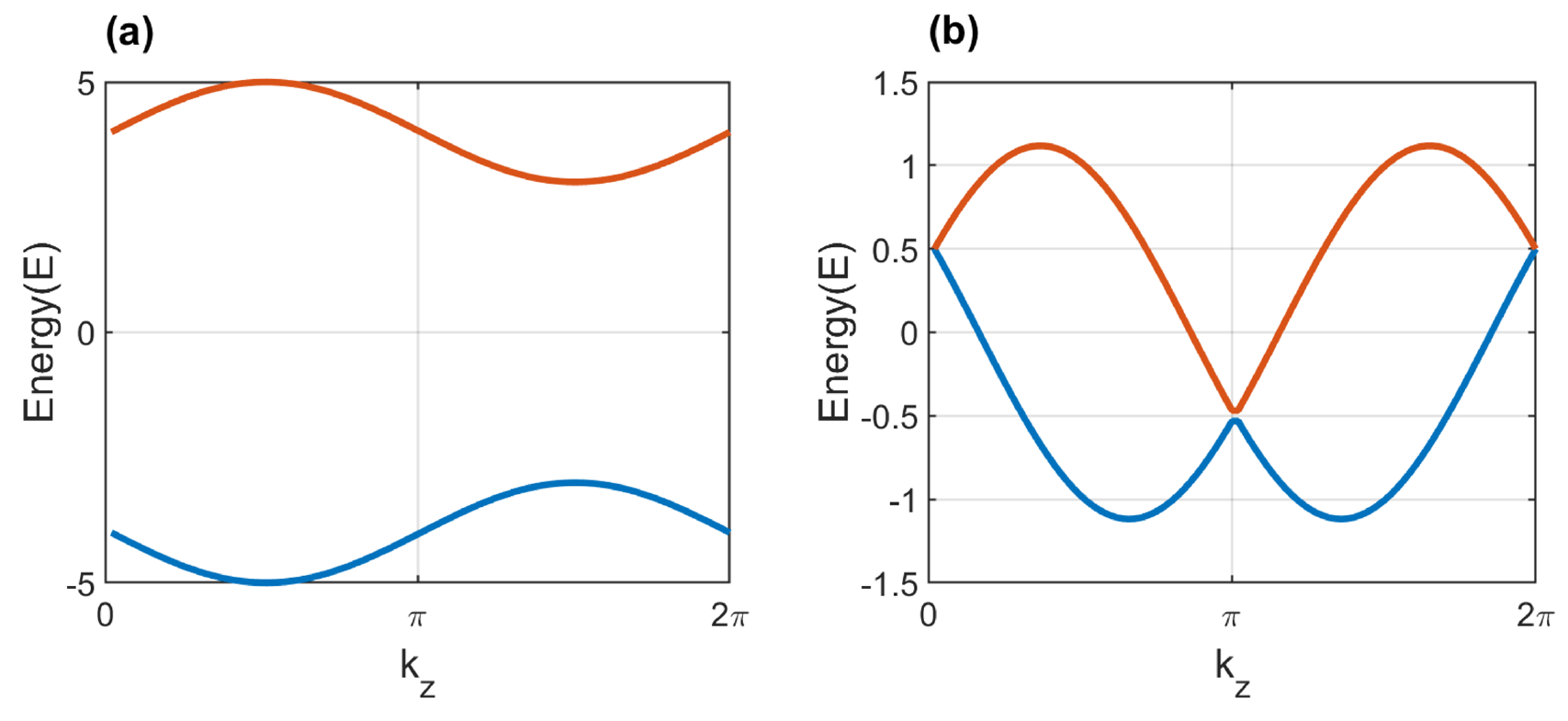}
 \caption{(a) Gapped energy spectrum of $H_0$ with parameters $(m_0,s) = (6,1)$ along $k_z$-axis at $(k_x, k_y) = (0, 0)$. (b) Energy spectrum alone $k_z$ of the Floquet-driven Hamiltonian with parameters $(k,s,\Omega, m_1) = (6,2,4,4)$. Gaps at $(0,0,0)$ and $(0,0,\pi)$ collapse so that  $(0,0,0)$ and $(0,0,\pi)$ become two Weyl nodes with chirality being $+1$ and $-1$ respectively.} \label{fig:Band_Structure}
\end{figure}

To analyze the  spectrum of $h_{\mathbf{k}}$,  let the eigenstate to the Floquet operator  $h_{\mathbf{k}} - i \partial_t$  be  $\ket{\psi_{a}(\mathbf{k},t)}$ so that  $(h_{\mathbf{k}} - i \partial_t) \ket{\psi_{a}(\mathbf{k},t)} = \epsilon_a \ket{\psi_{a}(\mathbf{k},t)}$ with $a=0,1$ being the band index and $\epsilon_a$ 
being the quasi-energy in the first Floquet Brillouin zone (FBZ).  $\ket{\psi_{a}(\mathbf{k},t)}$ can be expressed in its Fourier components as $\ket{\psi_{a}(\mathbf{k},t)} = e^{-i\epsilon_{a} t}\sum_n e^{-in\Omega t} \ket{\phi_a(\mathbf{k},n)}$, where $\epsilon_a + n \Omega$ is the quasi-energy associated with the nth component $\ket{\phi_a (\mathbf{k},n)}$.  In the frequency domain, $h_{\mathbf{k}}$ becomes the Floquet Hamiltonian $H_F$\cite{Floquet_2} and is given by
\begin{equation}
    H_F = \left[ \begin{matrix}
   H_{n,n},  &H_{n,n+1}, &... \\
   H_{n+1,n}  &H_{n+1,n+1}, &...\\
   ...  &H_{n+2,n+1}, &H_{n+2,n+2}
    \end{matrix}\right],
    \label{eq:Floqet_Ham}
\end{equation}
where 
$
    H_{m,n} = \int_0^{T_0} dt \, \left[ h_0({\mathbf{k}})+h_T({\mathbf{k}},t) \right] e^{-i(m-n)\Omega t} + \delta_{m,n} m\Omega$.
The corresponding band structure forms multi-pairs Weyl nodes in frequency domain. In Fig.\ref{fig:Band_Structure} (a),  we show the electronic structure of the topological trivial band before the Floquet driving is turned on; while in Fig.\ref{fig:Band_Structure} (b), the Floquet driving is turned on, the band in (a) is driven to a Weyl semimetallic phase. Here Fig.\ref{fig:Band_Structure}(b) shows the band structure in the first FBZ, in which two Weyl nodes appear at $(k_x, k_y, k_z) = (0,0,0), (0,0,\pi)$. These Weyl nodes are robust under parameter perturbation\cite{Weyl_nature}.  Furthermore, one can derive an effective Hamiltonian $h_{eff}$ to describe the Floquet driven spectrum. For this purpose, we first note that as the effective Hamiltonian that describes the $n$th Floquet band must be a $2 \times 2$ matrix. Hence it can be generally expressed as a summation of Pauli matrices as $ A \cdot I + B \sigma_x +C \sigma_y+D \sigma_z$ with the coefficients $A$, $B$, $C$, and $D$ being expanded in terms of $\sin k_i$ and $\cos k_i$ ($i=x$,$y$,$z$).  To get the form for these coefficients, we note that $n \Omega$ and $1/2 s \cos k_z$ are already in $H_{nn}$ in Eq.(4). Furthermore,  in $h_T({\mathbf{k}},t)$, $m_1$ acts as an ac magnetic field that induces coupling of the original electronic band in $H_0$ to the same energy band shifted in frequency ($\pm \Omega$).  Hence we expect  that a term $\Omega \sigma_z$ is present to reflect the energy shift. As $H_{n, n+1}$ and $H_{n,n-1}$ are the only non-vanishing coupling terms in Eq. (4) and they depend on $\sigma_z$ through the ac magnetic field term in Eq.(2), one expects to get combinations of $\sigma_x \sigma_z$ ($\propto \sigma_y$) or $\sigma_y \sigma_z$ ($\propto \sigma_x$) to the leading terms by treating $H_{n, n+1}$ and $H_{n,n-1}$ as perturbations to $H_{nn}$. We thus expect that  coefficients of $\sin k_x \sigma_x$ and $\sin k_y \sigma_y$ are corrected and introduce $\lambda_x$ and  $\lambda_y$ as the corrected coefficients that will be fixed by numerically-computed spectrum. From the above reasoning, we find that the resulting spectrum in  the $n$th FBZ can be described by a time-independent effective Hamiltonian as   
\begin{align}
   & h_{eff} ({\mathbf{k}}) = \lambda_x \sin k_x\sigma_x + \lambda_y \sin k_y\sigma_y + \sin k_z \sigma_z \nonumber \\
    &   + ( m_0 - \cos k_x - \cos k_y - \Omega )\sigma_z +\frac{s}{2} \cos k_z  \mathbf{1} + n\Omega  \mathbf{1}.
\end{align}
$h_{eff}$ has been tested by computing the spectrum of  Eq.(4) with large cutoff (the matrix involved is at least $1000 \times 1000$). The convergence of the resulting spectrum is also checked by changing the cutoff. $h_{eff}$ is found to accurately describe the computed spectrum.
To get the numerical values of $\lambda_x$ an $\lambda_y$, we fix $k_x$ and $k_y$ and compute the energy spectrum versus $m_1$. It is found that the spectrum is periodic in $m_1$ with period being around 16.
By fitting to numerical results, we find that $ \lambda_x = \lambda_y = 0.67 \cos(0.39  m_1 + 0.26)$. Note that as shown in Fig. 1(a), in the original Hamiltonian without the Floquet driving, the energy spectrum is in the range from $E=-10$ to $E=10$. Since $\Omega$ is the period of Floquet driving, which is also the unit cell in frequency space, to fill the unit cell in frequency, we only consider parameters of $\Omega$ and $m_1$ for values being less than $10$.  Furthermore, because the energy gap in the original Hamiltonian without the Floquet driving is about 5, to be able to hybridize the conduction and valence bands to yield the Weyl semi-metallic phase, it requires that $m_1$ and $\Omega$ have to be greater than $4$ when both terms in $h_T$ are considered. Hence overall speaking, the effective Hamiltonian and corresponding coefficients are tested and valid only in the range: $4 \leq m_1 \leq 10 $, $ 4 \leq \Omega \leq 10$. \\

 \begin{figure*}[t]
    \centering
    \includegraphics[width = 0.7\textwidth]{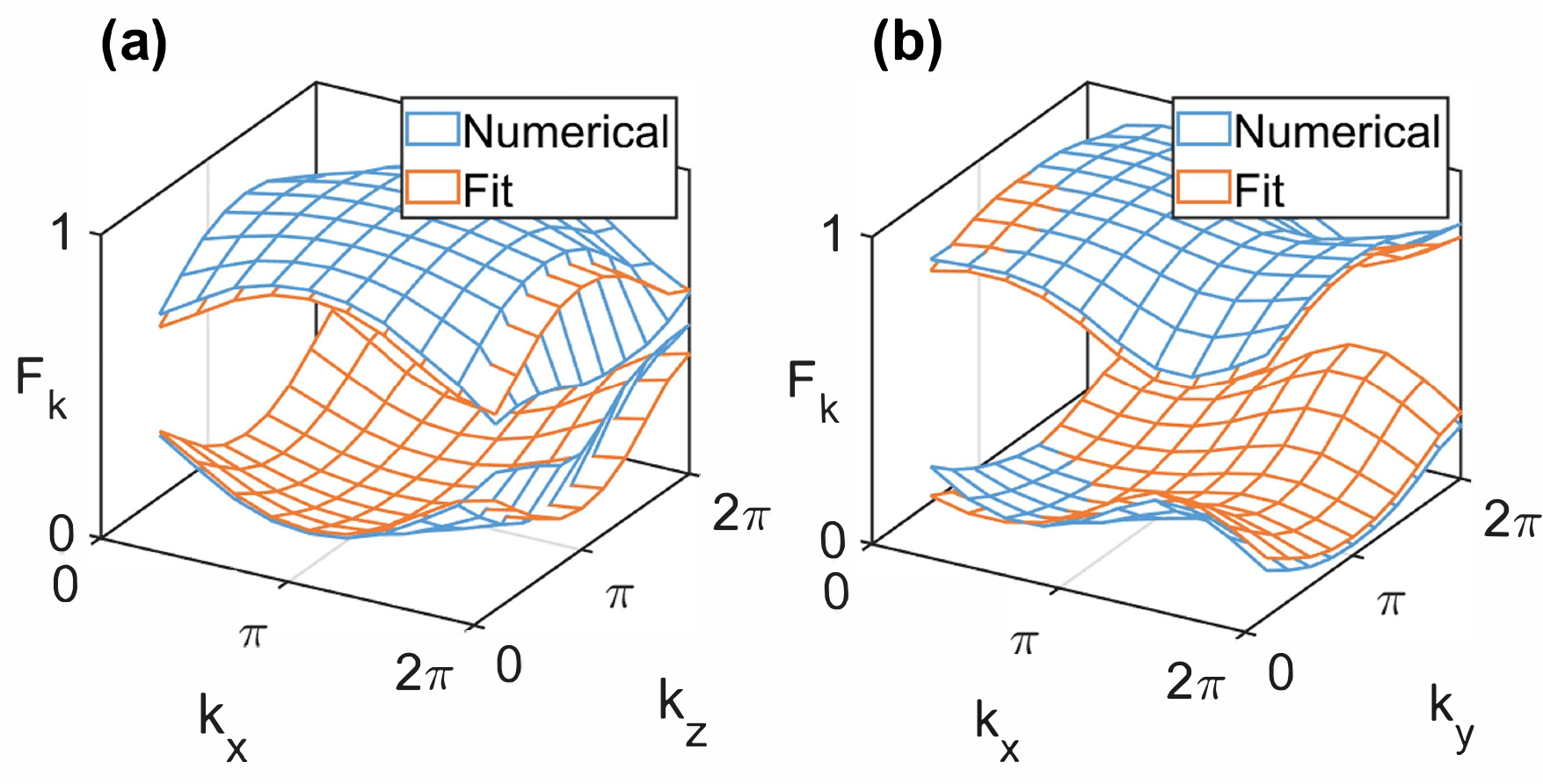}
    \caption{(a) The blue mesh is the numerical solution to Eq.(\ref{eq:Distribution}), plotting with arbitrarily chosen $k_y = \frac{4\pi}{5}$, and the orange mesh is the result fitted to the Fermi-Dirac distribution with fitted temperature being $T^* = 1.387$ and fitted chemical potential being $\mu^* = 0.016$. Here the momentum space is divided into $10 \times 10 \times 10$. The reservoir temperature is kept in $T = 0.001$ and $G_0 = 0.05 \Omega$.  Other parameters are the same as those used in Fig.\ref{fig:Band_Structure}(b). (b) Plot of $F_{\mathbf{k}}$ versus $(k_x,k_y)$ with $k_z = \frac{4\pi}{5}$}.
    \label{fig:Distribution}
\end{figure*}
\section{Floquet Occupation State} 
The Floquet driven system will heat up eventually if it is not a many-body localized system\cite{many_loc}. To prevent overheating, we couple the system to a bosonic reservoir\cite{ FloBlotz_Seetharam2019}. Following the discussion in Ref. (\onlinecite{FloBlotz_Seetharam2019}), since the Floquet energy gap in the Floquet-Weyl semimetal is zero, which is less than the finite bandwidth of the phonon reservoir $\Omega_D$, it ensures the cooling process in which the excited particle in the upper band can scatter back to the lower band by releasing a phonon. Hence a steady state can be established. In our setting, we further require the phonon band width $\Omega_D$ to be much smaller than the driven frequency, $\Omega_D \ll  \Omega$so that the Floquet-Umklapp process which may also heat up the system is suppressed. The bosonic reservoir is modeled by a phonon bath characterized by the Hamiltonian $H_{bath}$ in together with an electron-phonon interaction $H_{ep} $ as
\begin{align}
    & H_{ep} = \sum G(\mathbf{k}, \mathbf{k'},\mathbf{q}) (b^{\dagger}_{-\mathbf{q}} + b_{\mathbf{q}})f^{\dagger}_{\mathbf{k}\sigma}f_{\mathbf{k'}\sigma},\\
    &H_{bath} = \sum \hbar \omega_{\mathbf{q}} b^{\dagger}_{\mathbf{q}} b_{\mathbf{q}}.
\end{align}
Here $G(\mathbf{k},\mathbf{k'},\mathbf{q})$ characterizes the scattering of electrons in the Floquet system by phonons in the reservoir. When the screening of charges is weak, the electron-phonon scattering is dominated by long-range Coulomb scattering and is thus dominated in the forward direction\cite{forward}.  Due to low density of electronic states in Weyl semimetals, the screening of charges is weak. Hence we shall assume that the electron-phonon coupling is dominated in the forward direction and take $G = G_0 \, \delta(\mathbf{k} - \mathbf{k'}-\mathbf{q})$ with $G_0$ being a constant to simplify the calculation.  In Eq.(7),  $\hbar \omega_{\mathbf{q}}$ is the energy of the phonon in with momentum $\mathbf{q}$ . We shall adopt the continuum model of phonons such that $\omega_{\mathbf{q}} = C|\mathbf{q}|=Cq$. Here $C$ is the speed of sound whose relation to the Debye frequency cutoff $\Omega_D $ is given by $\sqrt{3}\pi C / a_p = \Omega_D$ with $a_p$ being the lattice constant for phonons. Following Ref. (\onlinecite{FloBlotz_Seetharam2019}),  the occupation number $F_{\mathbf{k} a} = \braket{f^{\dagger}_{\mathbf{k} a}f_{\mathbf{k} a}}$ with $a=0,1$ can be found by solving the Floquet-Boltzmann equation.  Starting from the quantum kinetic equation that $F_{\mathbf{k} a} $ obeys: $
    i\partial_t F_{\mathbf{k} a} = \braket{\Big[f^{\dagger}_{\mathbf{k} a}f_{\mathbf{k} a}, H_{ep}\Big]}$, one can derive the Floquet-Boltzmann equation through the perturbation theory.  In the derivation, it  requires the evaluation of terms such as $\braket{f^{\dagger}_{\mathbf{k}+\mathbf{q} a}f_{\mathbf{k} a}b_{\mathbf{q}}}$. By performing perturbative expansion and keeping the leading order terms in $G_0$, we find that $F_{\mathbf{k} a} $ satisfies the  Floquet-Boltzmann equation
 \begin{align}
     &  \partial_t F_{\mathbf{k} a}  = \sum_{a, b, \mathbf{q}} |G_0|^2  \delta(\epsilon_{a}(\mathbf{k}) - \epsilon_{b}(\mathbf{k}+\mathbf{q}) - \hbar \omega_{-\mathbf{q}} ) \nonumber\\
    &\times \left[ F_{\mathbf{k} a}(1 - F_{\mathbf{k}+\mathbf{q} b}) (1+N(\omega_{-\mathbf{q}})) -  F_{\mathbf{k}+\mathbf{q} b}(1 - F_{\mathbf{k} a}) N(\omega_{-\mathbf{q}}) \right] \nonumber \\
 & - \delta(-\epsilon_{a}(\mathbf{k}) + \epsilon_{b}(\mathbf{k}+\mathbf{q}) - \hbar\omega_{\mathbf{q}} )  \times \left[ F_{\mathbf{k} a} \longleftrightarrow F_{\mathbf{k}+\mathbf{q} b}, \right. \nonumber \\
& \left. \omega_{-\mathbf{q}} \longleftrightarrow  \omega_{\mathbf{q}}   \right] .\label{eq:Distribution}
 \end{align}
Here $a=0 $ or $1$ is the band index in FBZ and $N(\omega_{\mathbf{q}}) = (e^{\beta  \omega_{\mathbf{q}}} - 1)^{-1}$ is the phonon occupation number. In principle, the summation over momentum needs to include all Floquet-Umklapp processes. However, since we have taken  $\Omega_D \ll  \Omega$ to suppress the Floquet-Umklapp process, we shall focus on processes in the first Floquet Brillouin Zone. The delta function in Eq.(\ref{eq:Distribution}) enforces the momentum  and energy conservations during the electron-phonon scattering. Since as discussed in the above, forward scattering is dominated in the electron-phonon coupling, the integration of the phonon momentum $\bf{q}$ can be done in a similar way as what is done in 1D electronic systems coupling to 3D bosonic bath in Ref.(\onlinecite{FloBlotz_Seetharam2019}). Here we shall assume that the occupation numbers $F_{\mathbf{k}+\mathbf{q}  b}$ and $N(\omega_{\mathbf{q}})$ are dominated for $\bf{q}$ being the forward direction to $\bf{k}$ so that the summation $\sum_{a, b, \mathbf{q}}\delta(\epsilon_{a}(\mathbf{k}) - \epsilon_{b}(\mathbf{k}+\mathbf{q})- \hbar \omega_{\mathbf{q}} ) [\cdot]$ can be performed in the transverse direction $\bf{q}_{\perp}$ of $\bf{q}$ and replace $\sum_{\bf{q}}[\cdot]$ by $\sum_{\bf{q}}[\cdot]=\sum_{q\hat{k},\bf{q}_{\perp}}[\cdot]=\sum_{q\hat{k}} \int d \omega \bar{\rho} (q \hat{k},\omega) [\cdot]$ with $\bar{\rho} (q \hat{k}, \omega)$ being the partial density of state to the forward scattering direction $\hat{k}$~\cite{FloBlotz_Seetharam2019}. Furthermore, since there is only one solution to satisfy the momentum and energy conservation in most of the time, following Ref.(\onlinecite{FloBlotz_Seetharam2019}),  we shall take the approximation by treating $\rho$ as a constant for our reservoir model such that $\bar{\rho} =1$, $|\epsilon_{a}(\mathbf{k}) - \epsilon_{b}(\mathbf{k}+\mathbf{q}))| \leq \omega_q= Cq$; otherwise $\bar{\rho} =0$.
In the steady state, $\partial_t F_{\mathbf{k} a} = 0$, which when combined with Eq.(8), leads to a self-consistent equation for the occupation number $F_{\mathbf{k} a} = \braket{f^{\dagger}_{\mathbf{k} a}f_{\mathbf{k} a}}$.
By setting $\Omega_D = 0.1 \Omega$, we solve Eq.(\ref{eq:Distribution}) self-consistently. In Fig.~\ref{fig:Distribution}, we show typical numerical results of the occupation number $F_{\mathbf{k} a}$. We find that the  occupation number can be fitted by a Fermi-Dirac distribution with an effective temperature $T^*$ and an effective chemical potential $\mu^*$. In particular, after the system couples to the reservoir, the temperature increases to $T^* = 1.387$ and the chemical potential is found to be $\mu^* = 0.016$, while at the same time, the reservoir temperature  is kept at $T = 0.001$. The effective temperature depends on the coupling constant $G_0$ in $H_{ep}$ but we are not going to investigate the detailed dependence on each coupling constant but leave it for a future study\cite{FloBlotz_Seetharam2019}. However, the dependence of the effective temperature $T^*$ on the parameter $s$  and $\Omega$ that specifies the electronic structure is crucial and is listed in Table.\ref{EffT_Omg}. It is clear that the effective temperature is not sensitive to $s$. However, increasing driven frequency $\Omega$ increases the energy rate that is pumped into the system. Hence it leads to higher effective temperatures.

\begin{table}[!htp]
  \caption{Variation of effective temperature $T^*$ and effective chemical potential ${\mu}^*$ with respect to $s$ for $\Omega=4$ (upper table) and with respect to $\Omega$ for $s = 1$ (lower table). Note that  fillings of electrons for all cases shown  are at 
half-filling.}
  \begin{tabularx}{0.4\textwidth}{XXX}
    \hline\hline
    $s$ & $T^*$ & $\mu^*$ \\
    \hline
    $1$ & $1.387$ & $0.016$\\
    $2$ & $1.394$ & $0.024$\\
    
    \hline\hline
    $\Omega$ & $T^*$ & $\mu^*$ \\
    \hline
    $4$ & $1.387$ & $0.016$\\
    $5$ & $1.394$ & $0.024$\\
    $6$ & $1.721$ & $0.063$\\
    $7$ & $2.080$ & $0.120$\\
    $8$ & $1.972$ & $0.184$\\
    \hline\hline
  \end{tabularx}
    \label{EffT_Omg}
\end{table}

\begin{figure*}[!]
    \centering
    \includegraphics[width = 0.8\textwidth]{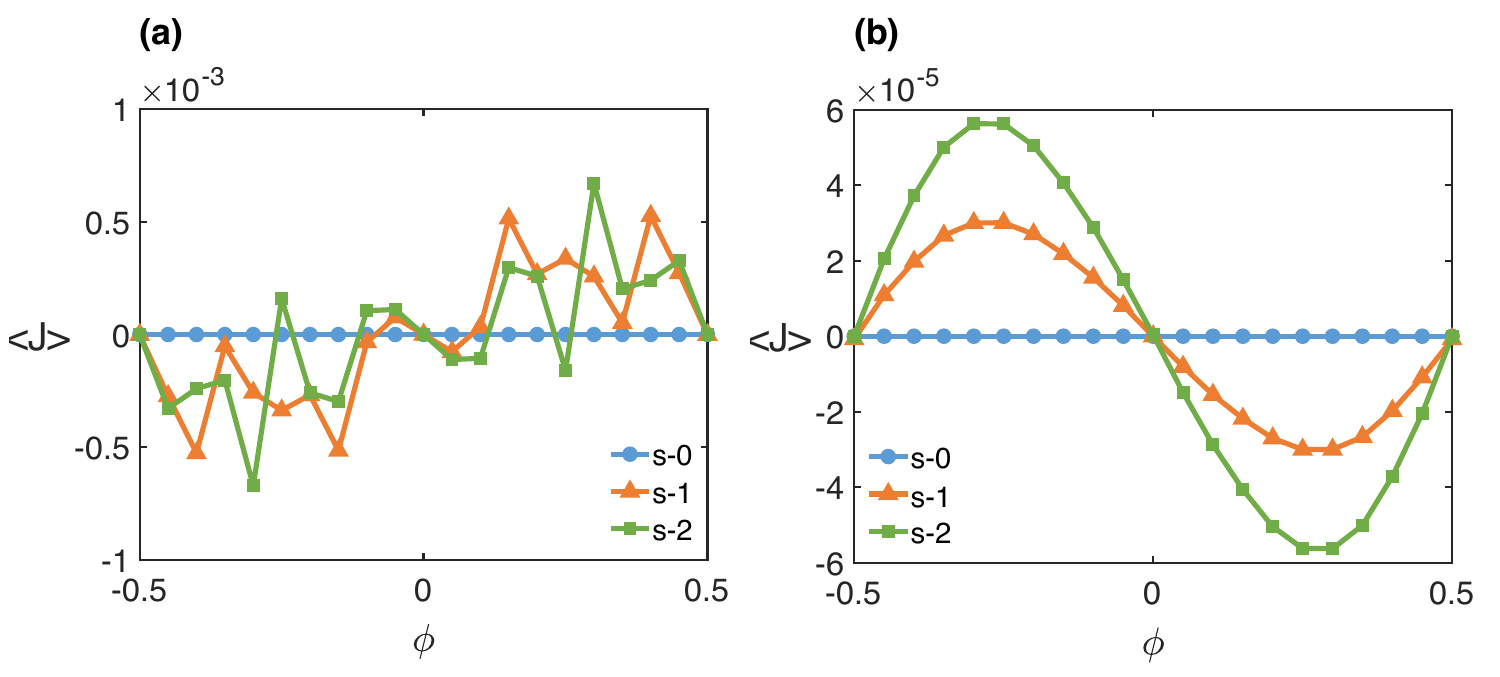}
    \caption{(a) Result of directly-computed CME current along $z$ direction in temperature $T=0$ for $s = 0,1,2$. Here Floquet bands with $n=0,-1,1$ are included and other parameters are the same as those adopted in Fig.\ref{fig:Band_Structure}. (b) CME current in $z$ direction is computed from the effective Hamiltonian with  the temperature being set at $T=1$}
    \label{fig:CME}
\end{figure*}
\section{Manifestation of Chiral Magnetic Current} 
Given non-vanishing occupation, the driven Floquet system can support non-vanishing current.  As indicated in the introduction section, the CME may exist in non-equilibrium systems. Therefore, we calculate the average current $\overline{\braket{J^i}}$ ($i = x, y,z$) over a period in the presence of magnetic field as
\begin{align}
  \overline{\braket{J^{i}}} \equiv \frac{1}{T_0}\int_0^{T_0} dt \braket{J^{i}(t)} &= \frac{1}{T_0}\int dt \int  d\mathbf{k} \sum_{a}  J^{i, a}_{\mathbf{k}}(t) F_{\mathbf{k} a}  (t), \label{eq:current_exp}
\end{align}
where $J^{i, a}_{\mathbf{k}}(t)= \sum_{m,n} \braket{\phi_{a} (\mathbf{k},m) | J^{i} (\mathbf{k},t) |\phi_{a} (\mathbf{k},n) }$ with $\mathbf{J} (\mathbf{k},t) = \partial h_{\mathbf{k}} /  \partial  \mathbf{k}$. In the frequency domain, we can express $ \braket{J^{i}}$ in the Floquet basis as
\begin{align}
   \braket{J^{i}}  &= \sum_{a} \Big[\sum_{m,n} J^{i, a}_{mn}(\mathbf{k})\Big] F_{\mathbf{k} a} (\epsilon_a).  \label{eq:current}
\end{align}
Here $J^{i, a}_{mn}$ is the matrix element of $J$ given by
\begin{equation} \label{eq:J}
     J^{i, a}_{mn}(\mathbf{k}) = e^{-i(n-m)\Omega t} \braket{\phi_{a} (\mathbf{k},m) | J^{i} (\mathbf{k},t) |\phi_{a} (\mathbf{k},n) }. 
\end{equation}
We then compute $\overline{\braket{J^{i}}}$ directly by choosing $i = z$ as the direction of the current and imposing a static magnetic field, $\mathbf{B} = (0, 0, B)$, which is also along $\hat{z}$ axis and passes through two Weyl points. In this case, $|\phi_{a} (\mathbf{k},n) \rangle $ in Eq.(\ref{eq:J}) will be replaced by eigenfunctions of Landau levels. By taking the vector potential in the Landau gauge with $\mathbf{A} = (-By, 0, 0)$, Landau levels are specified by $\mathbf{k}_{\perp}=(k_x,k_z)$ and index of lattice sites along $\hat{y}$ axis, $n_y$. Since any hopping amplitude $t_x$ along $\hat{x}$ direction is replaced by $ t_x e^{i2\pi \phi n_y}$ upon the Peierls substitution with $\phi = \frac{B}{\phi_0}$ and $\phi_0 = \frac{hc}{e}$ being the flux quanta, the $x$-component of the wave-vector, $k_x$, becomes $k_x + 2\pi n_y \phi$. Hence, the Hamiltonian can be written as
\begin{align} 
   H=\sum_{\mathbf{k}_{\perp}, n_y, m_y, a, b}  f^\dag_{\mathbf{k}_{\perp}, n_y, a} h^{a b}_{n_y,m_y} (\mathbf{k}_{\perp},t) f_{\mathbf{k}_{\perp},m_y, b} + h.c..  \label{eq:real_space}
\end{align}
Here $h_{n_y,m_y}(\mathbf{k}_{\perp},t) =  [  h_0 (k_x + 2\pi n_y \phi, k_z,t) - \sin k_y\sigma_y  + \cos k_y \sigma_z + h_T(k_x + 2\pi n_y \phi, k_z,t)] \delta_{n_y,m_y} +1/2(\sigma_z+i\sigma_y)\delta_{n_y-1,m_y} +1/2(\sigma_z-i\sigma_y)\delta_{n_y+1,m_y}$. After numerically solving eigenstates to the Hamiltonian in frequency space, we substitute the eigenfunctions into Eq.(\ref{eq:current})
and obtain $\braket{J^i}$ in uniform magnetic fields. The numerical result is shown in Fig.\ref{fig:CME}(a), which clearly exhibits non-vanishing CME current at temperature $T=0$ for non-vanishing $s$. For a comparison,  in Fig. \ref{fig:CME}(b), we show the CME current obtained in the corresponding effective Hamiltonian with the temperature being set at $T=1$. Clearly,  it confirms that non-vanishing CME currents arise for non-vanishing $s$. For both temperatures ($T=0$ and $T=1$), when $s=0$, two Weyl points collapse into one, the CME current vanishes.  Furthermore, the CME current changes sign as one goes from $T=0$ to $T^* \approx 1.394 $ and almost decreases to zero at $T^* \approx 1.394$. The numerical computation in Fig.\ref{fig:CME}(a) can be further extended to finite temperatures and it shows similar results as those shown in Fig.\ref{fig:CME}(b). We note in passing that while the above non-vanishing CME current is computed by using the simulated Fermi-Dirac-like population distribution, the exact form of population is not essential for non-vanishing CME current to survive. In fact, by using Gaussian distributions, we still obtain non-vanishing  CME current. It shows that as long as contributions from two Weyl points are not the same, finite CME current will survive.

The direct computation of the CME current shown in the above is reliable only for finite and large magnetic fields, in which the lattice effect is important so that the energy spectrum exhibiting features of the Hofstadter spectrum.
For small magnetic fields and when $B$ approaches zero, Landau levels are dense in energy with total number approaching infinity. In this case, it is more reliable to compute 
the CME current as a linear response through the relation $ \braket{\vec{J}} = \alpha \vec{B}$.  Here the linear response is $\alpha$, defined as the CME coefficient. To obtain $\alpha$,  we first note that the linear response of the average current can be generally expressed as\cite{mou1997, mou_2018}
\begin{align}
    \braket{J_i(\mathbf{q},\omega)} &= \Pi_{ij} (\mathbf{q},\omega) \, A^j(\mathbf{q},\omega)  \equiv  i \alpha \epsilon^{ijk} q_k A^j(\mathbf{q},\omega). \label{eq:retard_current}
\end{align}
Here $\Pi_{ij}$ is the retarded current-current correlation function\cite{mou1997}. By taking $\mathbf{q}=q \hat{k}$ in Eq.(\ref{eq:retard_current}) and choosing $z$ as the direction of the current, $\alpha$ can be determined by the anti-symmetric part of $\Pi$ as
$ \alpha = -\frac{i}{2q}( \Pi_{ij} - \Pi_{ji})$\cite{mou_2018}. In the  Floquet basis,  $\Pi_{ij}$ can be expressed as
\begin{align}
    \Pi_{ij} (\mathbf{q};\omega) &= \frac{i}{V\Tr[\rho (t)]} \sum_{\mathbf{k}} \sum_{ab} \sum_{mn} \nonumber \\
    &\times  \rho^a \frac{\Tr[J^i P^b_n(\mathbf{k}+\mathbf{q}) J^j P^a_m(\mathbf{k}) ]}{\omega + \epsilon_a^m(k) - \epsilon_b^n(\mathbf{k}+\mathbf{q}) + i\delta}. 
\end{align}
Here $\rho (t)$ is the density matrix with its diagonal component being denoted by $\rho^a$. $V$ is the volume of the system.  $ P^a_m(\mathbf{k})$ is the projection operator $\ket{\phi_a(m,\mathbf{k})}\bra{\phi_a(m,\mathbf{k})}$. $\epsilon_a^m = \epsilon_a + m \Omega$ is the energy of band $a$ with Floquet index $m$. In the dc limit: $\mathbf{q} \longrightarrow 0 $ and $\omega \longrightarrow 0 $,  by expanding  $ \Pi_{ij} (\mathbf{q};\omega)$ in $\mathbf{q}$ and using the identity
$(\partial_{\mathbf{q}}  \bra{ \phi_{b} (\mathbf{k}, n)})  \ket{\phi_{c} (\mathbf{k}, p)} = \braket{\phi_{b} (\mathbf{k}, n)|J^q|\phi_{c} (\mathbf{k}, p)}/(\epsilon_n^b -\epsilon_p^c)$  when
$(b,n) \neq (c,p)$, the CME coefficient $\alpha$ can be expressed as
\begin{align}
    \alpha &=  \frac{i}{V\Tr[\rho(t)]} \sum_{\mathbf{k}} \sum_{abc} \sum_{mnp} \nonumber \\
    &\times  \rho^a \frac{\Tr[J^i P^c_p J^k P^a_m J^j P^b_n]}{(\epsilon_a^m - \epsilon_b^n)(\epsilon_c^p - \epsilon_b^n)} + \frac{\Tr[J^i P^b_n J^k P^c_p J^j P^a_m ]}{(\epsilon_a^m - \epsilon_b^n)(\epsilon_c^p - \epsilon_b^n)} \nonumber \\
    & - (i\longleftrightarrow j) \, \mbox{with} \, (a,m) \neq (b,n)\, , (c,p) \neq (b,n) \, . \label{eq:alpha}
\end{align}
Here $i$, $j$, and $k$ are three cyclic indices for $x$, $y$ and $z$. $(a,b,c) \in \{0,1\}$ are the band indices. $(m,n,p)$ are the Floquet indices from $-\infty$ to $\infty$. Note that due to the constraint on band indices, the only possible cases are $(a,b,c) = (0,1,0), (1,0,1)$.  In Fig.~{\ref{fig:Alpha-S}, we show numerically computed $\alpha$ versus $s$.  We see that $\alpha$ generally does not vanish and decreases as $s$ increases.  The non-vanishing $\alpha$ for $s \neq 0$ further confirms the existence of non-vanishing CME current.  In together with the direct computation of  CME current as shown in Fig.~3(a) for finite and large magnetic fields, we conclude that  in the presence of a static magnetic field, the steady Floquet-Weyl semimetal supports macroscopic chiral magnetic current. 
\begin{figure}[tp]
    \centering
    \includegraphics[width = 8cm]{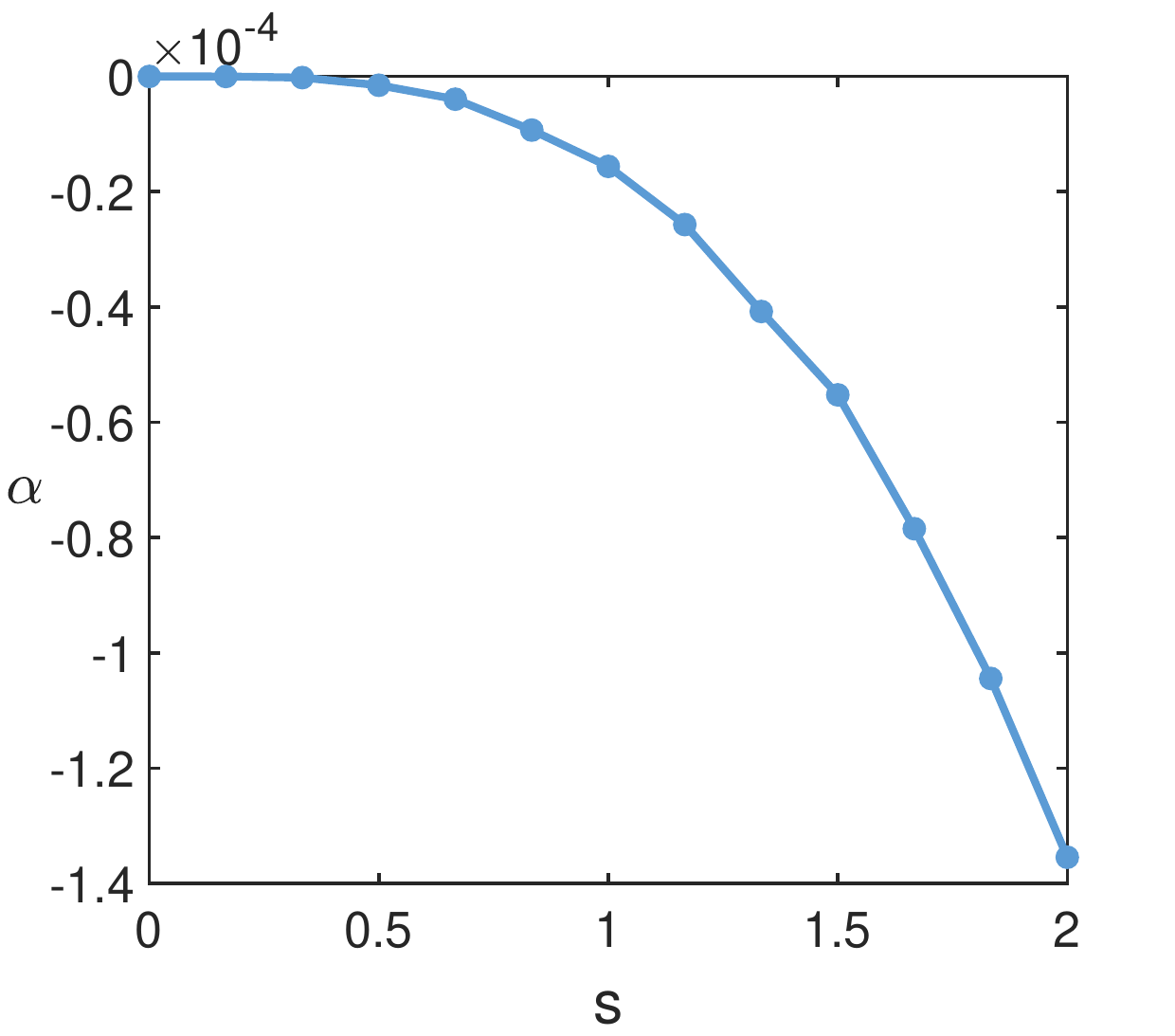}
    \caption{CME coefficient $\alpha$ versus $s$ calculated from Eq.(\ref{eq:alpha}) at zero temperature.}
    \label{fig:Alpha-S}
\end{figure}

\section{Discussion and conclusion}
In comparison to the equilibrium condensed-matter systems in which the electric current can be expressed as the total derivative of band energy dispersion, 
the expression of the electric current in Floquet systems is given by Eq.(\ref{eq:current}) and can not be expressed as a total derivative of the band energy. The  mechanism for non-vanishing CME current in Floquet systems is best illustrated by considering the current expression in Eq.(\ref{eq:current}) and Eq.(\ref{eq:alpha}).  Due to dynamical transitions between different Floquet bands in frequency space, the current operator is not diagonal in Floquet index. As a result, the off-diagonal matrix element of current operator in Eq.(\ref{eq:current}) can not be reduced to the total derivative of energy eigenvalues with respect to $\mathbf{k}$, which is shown to vanish in equilibrium condensed-matter systems\cite{nogo}. Indeed, from Eq.(\ref{eq:alpha}), it is clear that dynamical transitions from Floquet bands in different frequencies to the same Floquet band give rise to non-vanishing CME coefficient.

In conclusion, we have shown that in a time-periodic potential,  an insulator can be driven to become a Floquet-Weyl semimetal with two Weyl points separated in energy. Furthermore, by coupling the resulting Floquet-Weyl semimetallic band to a phonon heat bath, we show that the electronic populations can be controlled so that the steady state is characterized by the Fermi-Dirac-like distribution. As a result, even in the presence of a static magnetic field, we find that the resulting steady Floquet-Weyl semimetal supports non-vanishing chiral magnetic current so that the chiral magnetic effect is realized in the condensed matter system. 

Our results indicate that dynamical transitions between different Floquet bands in frequency space is the key to break the no-go theorem of CME current in equilibrium condensed-matter systems. As the CME current is a macroscopic persistent current, our results open the door to realize novel macroscopic persistent current in condensed matter systems.  

\begin{acknowledgments}

This work was supported by the National  Science and Technology Council (NSTC), Taiwan. We also acknowledge support from the Center for Quantum Science and Technology (CQST) within the framework of the Higher Education Sprout Project by the Ministry of Education (MOE) in Taiwan.

\end{acknowledgments}


\begin{thebibliography}{99}
\bibitem{persist} J. File J and R. G. Mills, Phys. Rev. Lett. {\bf 10}, 93 (1963).
\bibitem{persist_2} L.P. L\'evy, G. Dolan, J. Dunsmuir, and H. Bouchiat, Phys. Rev. Lett. {\bf 64}, 2074 (1990).
\bibitem{graphene} K. S.Novoselov,A.K. Geim, S.V. Morozov, D. Jiang,Y. Zhang,
S. V. Dubonos, I. V. Grigorieva, and A. A. Firsov, Science {\bf 306}, 666 (2004); 
A. H. Castro Neto, F. A. H. Guinea, N. M. R. Peres, K. S. Novoselov, and A. K. Geim, Rev. Mod. Phys. {\bf 81}, 109 (2009).
Novoselov, and A. K. Geim, Rev. Mod. Phys. 81, 109 (2009).
\bibitem{Kane2005_7} C. L. Kane and E. J. Mele, Phys. Rev. Lett. {\bf 95},146802 (2005); L. Fu, C. L. Kane, and E. J. Mele, Phys. Rev. Lett. {\bf 98}, 106803 (2007).
\bibitem{Zhang2012} B. Yan and S.-C. Zhang, Reports on Progress in Physics {\bf 75}, 096501 (2012).
\bibitem{Liang2008} X. L. Qi, T. L. Hughes, and S.-C. Zhang, Phys. Rev. {\bf B 78}, 195424 (2008).
\bibitem{Yan2017} B. Yan and C. Felser, Annual Review of Condensed Matter Physics {\bf 8}, 337 (2017).
\bibitem{CME} For recent reviews, see P. Hosur and X. L. Qi, C. R. Phys. {\bf 14}, 857
(2013); D. E. Kharzeev, Prog. Part. Nucl. Phys. {\bf 75}, 133 (2014).
\bibitem{Dirac1} S. Murakami, New J. Phys. {\bf 9}, 356 (2007).
\bibitem{mou1} P.-H. Chou, L.-J. Zhai, C.-H. Chung, C.-Y. Mou, and T.-K. Lee, Phys. Rev. Lett. {\bf 116}, 177002 (2016).
\bibitem{Dirac2} Z. K. Liu et al., Science {\bf 343}, 864 (2014).
\bibitem{Dirac3} S. Borisenko, Q. Gibson, D. Evtushinsky, V. Zabolotnyy, B. Buchner, and R. J. Cava, Phys. Rev. Lett. {\bf 113}, 027603 (2014).
\bibitem{Dirac4} M. Neupane, Nat. Commun. {\bf 5}, 3786 (2014).
\bibitem{Weyl1} S.-Y. Xu et al., Science {\bf 349}, 613 (2015).
\bibitem{Weyl2} B.Q. Lv, H. M. Weng, B. B. Fu, X. P. Wang,H.Miao, J. Ma, P.Richard, X. C. Huang, L. X. Zhao, G. F. Chen, Z. Fang, X. Dai,
T. Qian, and H. Ding, Phys. Rev. X {\bf 5}, 031013 (2015).
\bibitem{Weyl3} L.-J. Zhai, P.-H. Chou, and C.-Y. Mou, Phys. Rev. {\bf B 94}, 125135 (2016).
\bibitem{Weyl4} A. A. Zyuzin andA.A. Burkov,Phys. Rev. {\bf B 86}, 115133 (2012).
\bibitem{Nielsen} H. B. Nielsen and M. Ninomiya, Nucl. Phys. {\bf B 185}, 20 (1981).
\bibitem{nogo} M. M. Vazifeh and M. Franz, Phys. Rev. Lett. {\bf 111}, 027201 (2013).
\bibitem{w_01} P. Baireuther, J. A. Hutasoit, J. Tworzydlo, and C. W. J. Beenakker, New J. Phys. {\bf 18}, 045009 (2016).
\bibitem{w_02} S. Zhong, J. E. Moore, and I. Souza,Phys. Rev. Lett. {\bf 116}, 077201 (2016).
\bibitem{w_03}  P. Goswami, G. Sharma, and S. Tewari, Phys. Rev. {\bf B 92}, 161110(R) (2015).
\bibitem{w_04}  A. Sekine and K. Nomura, Phys. Rev. Lett. {\bf 116}, 096401 (2016).
\bibitem{w_05} J. Ma and D. A. Pesin, Phys. Rev. Lett. {\bf 118}, 107401 (2017).
\bibitem{Floquet_1} G. Floquet, Annales scientifiques de l'\'Ecole normale sup\'erieure, {\bf 12}, 47 (1883).
\bibitem{Floquet_2} A. G\'omez-Le\'on and G. Platero, Phys. Rev. Lett. {\bf  110}, 200403 (2013).
\bibitem{Floquet_top_1} X.-X. Zhang, T. T. Ong, and N. Nagaosa, Phys. Rev. {\bf B 94}, 235137 (2016).
\bibitem{Floquet_top_2} Yufei Zhu, Tao Qin, Xinxin Yang, Gao Xianlong, and Zhaoxin Liang, Phys. Rev. Research 2, 033045 (2020).
\bibitem{Floquet_top_3} Leda Bucciantini, Sthitadhi Roy, Sota Kitamura, and Takashi Oka, Phys. Rev. {\bf B 96}, 041126(R) (2017).
\bibitem{Floquet_top_4} Jin-Yu Zou and Bang-Gui Liu, Phys. Rev. {\bf B 93}, 205435 (2016).
\bibitem{Floquet_CME_1} Shu Ebihara, Kenji Fukushima, and Takashi Oka, Phys. Rev. {\bf B 93}, 155107 (2016). 
\bibitem{Floquet_CME_2}  Katsuhisa Taguchi, Tatsushi Imaeda, Masatoshi Sato, and Yukio Tanaka, Phys. Rev. {\bf B 93}, 201202(R) (2016).
\bibitem{Floquet_CME_3} Sho Higashikawa, Masaya Nakagawa, and Masahito Ueda, Phys. Rev. Lett. {\bf 123}, 066403 (2019).
\bibitem{Floquet_CME_4} Xiao-Qi Sun, Meng Xiao, Tomas Bzdusek, Shou-Cheng Zhang, and Shanhui Fan, Phys. Rev. Lett. {\bf 121}, 196401 (2018).
\bibitem{Burkov2011} A. A. Burkovv and L. Balents, Phys. Rev. Lett. {\bf 107}, 127205 (2011).
\bibitem{Weyl_nature} A similar stable Floquet-Weyl semimetallic phase was also realized in systems driven by laser. See H. Hubener, M. A. Sentef, Umberto De Giovannini, A. F. Kemper, and A. Rubio, Nat. Commun. {\bf 8}, 13940 (2017).
\bibitem{many_loc}  K.S.C  Decker, C. Karrasch, J. Eisert, and D. M.  Kennes, Phys. Rev. Lett. 124, 190601 (2020).
\bibitem{FloBlotz_Seetharam2019} K. I. Seetharam, C.-E. Bardyn, N. H. Lindner, M. S. Rudner, and G. Refael, Phys. Rev. {\bf B 99}, 014307 (2019).
\bibitem{forward} O. V. Dolgov, O. V. Danylenko, M. L. Kulic, and V. Oudovenko, Int. J. Mod Phys. B 12, 3083 (1998).
Lee, et al., Science {\bf 349}, 613 (2015).
\bibitem{mou1997} C. Y. Mou, Phys. Rev. B {\bf 55}, R3378(R) (1997).
\bibitem{mou_2018} Y. T. Lin,  L. J. Zhai, and C. Y. Mou, Phys. Rev. B {\bf 97}, 245121 (2018).
\end{thebibliography}

\end{document}